\documentclass{caosp}

\usepackage{graphicx}

\articleNo{180} \pubyear{2009} \volume{39} \volnumber{1}
\firstpage{34} \lastpage{42} \received{October 29, 2008}
\accepted{February 2, 2009}

\begin{document}

\hauthor{P.\,Vere\v{s}, J.\,Budaj, J.\,Vil\'{a}gi, A.\,Gal\'{a}d and
L.\,Korno\v{s}}

\title{Relative photometry of transiting exoplanet COROT-Exo-2b}

\author{P.\,Vere\v{s} \inst{1}
 \and
 J.\,Budaj \inst{2}
 \and
 J.\,Vil\'{a}gi \inst{1}
 \and
 A.\,Gal\'{a}d \inst{1}
 \and
 L.\,Korno\v{s} \inst{1}}

\institute{Department of Astronomy, Physics of the Earth and
Meteorology, Faculty of Mathemtacics, Physics and Informatics,
Comenius University, 842\,48 Bratislava, The Slovak Republic
\email{veres@fmph.uniba.sk}
\and
\lomnica}

\date{October 29, 2008}

\maketitle

\begin{abstract}
Relative CCD photometry of the extrasolar planet COROT-Exo-2b transiting
in front of its parent star was carried out at the Astronomical and
Geophysical Observatory of Comenius University at Modra (AGO).
Physical and orbital parameters were determined and compared with the
previous published data.
\keywords{extrasolar planet -- relative photometry}
\end{abstract}

%
\section{Introduction}
\label{intr} The idea of planets orbiting another stars appeared
through philosophic ideas of ancient Greeks, centuries later in
book by Huygens (Huygens, 1698) and finally since the middle of
the 20th century. Despite reasonable methods observers were not
successful (e.g. Strand, 1943; Reuyl and Holmberg, 1943; van de
Kamp, 1963; Harrington {\it et al.}, 1983; McCarthy {\it et al.},
1985). Recent development of observation techniques and
instrumentation finally led to the first discovery of a planet
orbiting a star - surprisingly a pulsar (Wolszczan and Frail,
1992) by a pulse timing method. Three years later the first planet
orbiting a sun-like star 51 Pegasi was discovered (Mayor and
Queloz, 1995) by radial velocity method. Yet, first planets were
strange in comparison to those in our Solar system - usually with
their mass several times greater than the mass of Jupiter and with
orbital periods of only several days, heavily irradiated by their
parent star ("hot Jupiters"). 333 objects classified as extrasolar
planets (exoplanets) were known before December 10, 2008 ({\it The
Extrasolar Planets Encyclopaedia, 2008}), most of them detected
thanks to precise spectroscopy (a radial velocity method) and
photometry (a transit method), 31 of them in multiple systems.
Currently, the object is considered an exoplanet if its $M\sin i$
is less than $13\,M_{jup}$ (Jupiter masses, approx. a boundary
between planets and brown dwarfs, lighter objects never go through
a Deuterium burning stage, e.g. Burrows {\it et al.}, 1997),
orbiting s star or multiple star systems. Several ground-based
(HAT - Bakos {\it et al.}, 2002; OGLE - Udalski {\it et al.},
2002; TrES - Alonso {\it et al.}, 2004, WASP - Pollaco {\it et
al.}, 2006) and the first space-based exoplanet survey (COROT) are
carried out and new discoveries increase rapidly. Less massive
planets are discovered. Some data become inaccurate because the
errors spread in time from published dates. Photometry of a
transiting exoplanet could reveal the transit time, depth of
transit, duration of transit, revolution period, radius of planet
and the inclination of the orbit quite precisely and the method is
applicable also for a small astronomical telescope. We present the
first successful attempt to detect an extrasolar planet transit
from Slovakia, calculate physical parameters of the system and
update the transit time and reduce errors.
\section{COROT-Exo-2b} COnvection ROtation and planetary Transits
(COROT) is a space mission telescope led by ESA and French Space
Agency, launched on December 27, 2006. The\,27 cm telescope is
aimed for two main goals: discovery of short period transiting
exoplanets and asteroseismology. The advantage of the mission lies
in ability to scan the same field of view for a long period of
time (weeks) without an interruption and perform almost real time
photometric analysis of tens of thousands stars. Also its
space-based position allows to acquire high precision photometry
of the order of $10^{-3}$ magnitude in one exposure. COROT-EXO-2b
was the second exoplanet discovered by this mission (Alonso {\it
et al.}, 2008). The "hot Jupiter" like planet is rotating around
K0V type star GSC 00465-01282 at $930$\,ly distance (Bouchy {\it
et al.}, 2008) in Aquila constellation. Physical properties of the
parent star and planet are given in Table 1. It is assumed that
the planet originated much farther from the parent star and has
overcome substantial orbital evolution. Strong tidal effects from
its parent star might have affected the physical and orbital
properties of the planet (Jackson {\it et al.}, 2008, Pont, 2008).
The detailed lightcurve analysis of the star confirmed the
existence of stellar spots and facular fields, also the stellar
rotation rate is relatively high (4.5\,days). The star is most
likely more active than our Sun. The oscillation of spotted
photospheric regions is equal to 10 synodic periods of the planet
and suggests the magnetic interaction between the exoplanet and
parent star (Lanza {\it et al.}, 2008). Alonso et al. (2008b)
detected a tentative secondary eclipse at the level of
$5.5$x$10^{-5}$. They also studied the O-C diagram and concluded that
there are no periodic variations in the O-C residuals larger than
~10\,s. Many observations of COROT-Exo-2b transits with the basic
parameters derived (such as the transit duration, epoch and depth) can
be found in the Exoplanet Transit Database.

\section{Observation and data reduction}
The observation of COROT-Exo-2b was performed from the AGO Modra (Code
118) by the 60cm main Cassegrain type telescope equipped with a CCD
camera Apogee Ap8 in primary focus. We observed during 4 nights,
beginning on August 27, September 17, October 1 and 15, whereby
the object had to reach the minimal altitude 30 degrees above the
horizon. Good photometric conditions were needed since the
magnitude drop during the transit was expected to be only
0.032\,{\it mag} in the Johnson R-filter according to data by Alonso
et al., 2008. We chose the relative photometry as a method,
exposing 30 seconds each image in the R-filter. Data reduction was
made with the MaxIm DL and CCD image analysis software by Vil\'{a}gi,
2007. The brightness of the parent star COROT-Exo-2 was compared
with 8 stars in the field of view of comparable brightness and
spectral type with no lightcurve variability (not a variable star
according to used catalogues GSC and USNO A2, no variability
observed). The overall standard photometric error for comparison
stars during the observation reached 0.003\,{\it mag} on August
27, 0.009\,{\it mag} on September 17, 0.015\,{\it mag} on October
1 and 0.02\,{\it mag} on October 15. As seen on quality of
photometry, the weather and seeing conditions during the third and
fourth night did not provide data with sufficient photometric
precision, therefore data from the first two nights were used for
lightcurve analysis.

\section{Data analysis}

Reduction of data from the first night immediately showed the
decrease and increase of the parent star brightness among the
quasi stable brightness of comparison stars. Only during this
night we detected the entire progress of transit and short periods
before the first and after the fourth contact between the planet
and star disks. To determine the beginning and the end of the
transit we used the AVE software (Barbera, 2000) which used the Kwee - van Woerden method (Kwee and van Woerden, 1956). Then the
middle of the transit (transit epoch) and transit duration were
derived. The second night's observation did not cover the whole
transit and the reduced data were therefore used for lightcurve
shape improvement only. The transit epoch from the first night
with the orbital period of the planet (Alonso {\it et al.}, 2008) were
used for linking the data from the second night of observation. To
find selected physical and orbital parameters of COROT-Exo-2b we
had to find a model lightcurve which would fit the measured data.

\begin{table}[t]
\small
\begin{center}
\caption{Physical properties and orbital parameters of extrasolar
planet COROT-Exo-2b and its parent star, left - previously
published values, right - our results.} \label{t1}
\begin{tabular}{llll}
\hline\hline
star (GSC 00465-01282) & Alonso {\it et al.}, 2008 &\\
\hline
$\alpha (2000)$        & 19\,h 27\,m 06.494\,s  &            \\
$\delta (2000)$        & +$01^{\circ}$\,23'\,\,   01.17''     &            \\
Spectral Type          & K0V                  &            \\
Apparent Magnitude (V) &  12.57               &            \\
Distance               & 930\,ly              &            \\
Mass                   & 0.97 $\pm 0.06\,M_{\sun}$ &       \\
Effective Temperature  & $5625 \pm 120\,K$    &            \\
Radius                 & $0.902 \pm 0.018\,R_{\sun}$ &     \\
\hline\hline
planet (COROT-Exo-2b) & Alonso {\it et al.}, 2008 & our results\\
\hline
Mass & $3.31 \pm 0.16\,M_{jup}$ & \\
Semimajor axis & $0.0281 \pm 0.0009\,AU$ &\\
Orbital period & $1.7429964 \pm 0.0000017\,day$ &\\
Eccentricity & 0 &\\
Radius &$1.465 \pm 0.029\,R_{jup}$& $ 1.318 \pm 0.158\,R_{jup}$\\
Transit epoch (HJD) & $ 2454706.4016 \pm 0.03766 $ &  $2454706.4041
\pm 0.0030$\\
Inclination & $87.84 \pm 0.10$&$ 87.88 \pm 0.15$\\
Transit duration & $2.28 \pm 0.06$ $hours$ & $2.24 \pm 0.15$ $hours$    \\
Transit depth & 0.032 $\pm$ 0.002 & 0.030 $\pm$ 0.007 \\
 \hline\hline
\end{tabular}
\end{center}
\end{table}

\begin{figure}[htp]
\centerline{\includegraphics[height=110mm,width=120mm]{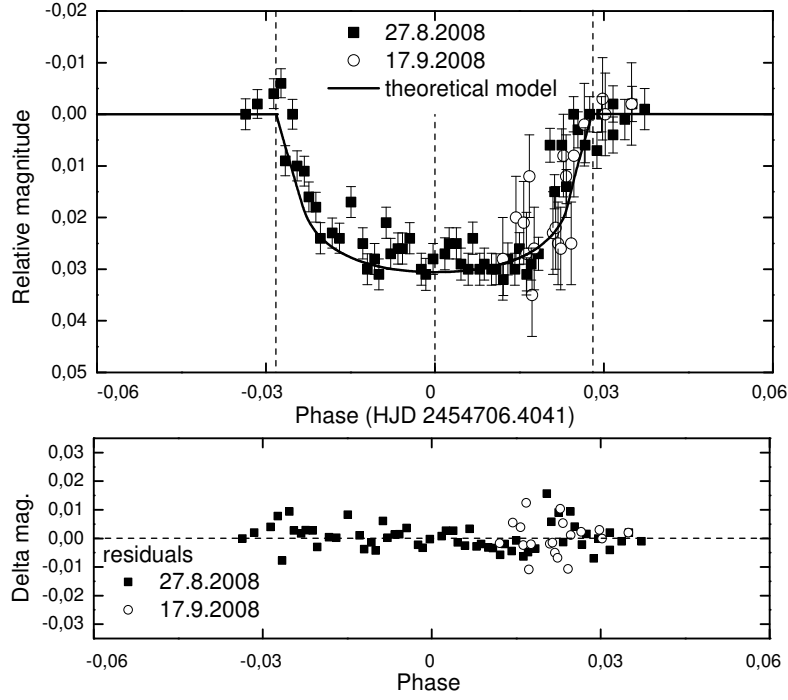}}
\caption{Photometric curve (top) and residuals from our
theoretical model (bottom) of transiting extrasolar planet
COROT-Exo-2b, data from second night were given to a phase
according to first night data, the beginning, center and the end
of the transit is marked.} \label{Figure 1}
\end{figure}

\begin{figure}[htp]
\centerline{\includegraphics[height=100mm,width=120mm]{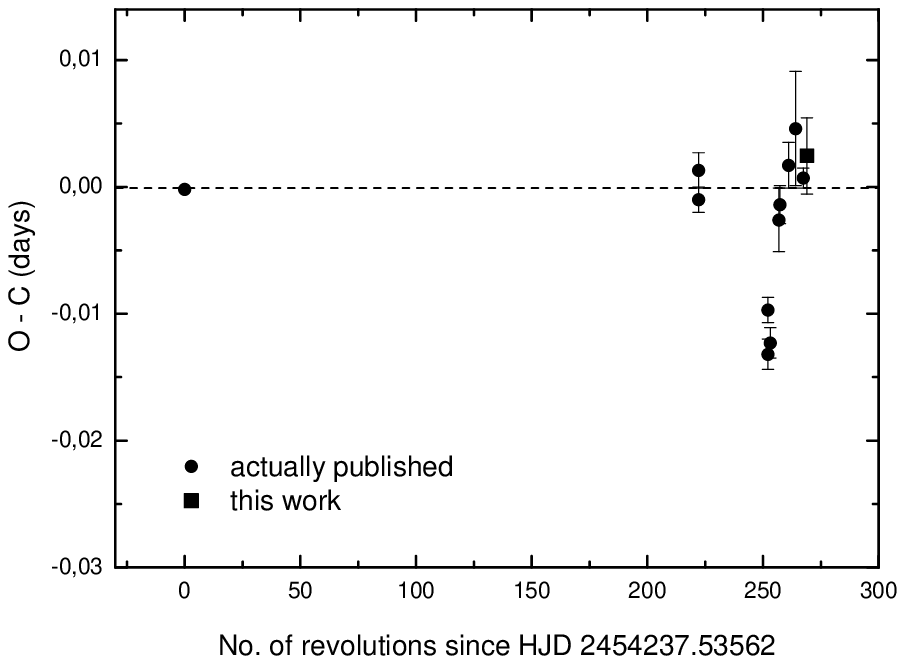}}
\caption{The O - C diagram of all actually published observations. It
shows how the observed minus calculated transit epoch differs from the time of observation. X-axis represents the number of planet
revolutions around the parent star since the discovery.} \label{Figure 2}
\end{figure}

The real shape of the photometric curve depends on the inclination of
the orbit {\it i}, orbital period {\it P}, planet semimajor axis
{\it a}, diameter of the planet {\it $R_{P}$} and the star {\it
$R_{*}$} (Cassen {\it et al.}, 2006). We consider that the stellar
disk is not uniformly bright and hat the
orbital eccentricity {\it e} is zero and {\it P}, {\it a} are
known (Alonso {\it et al.}, 2008). We used a quadratic expression
for limb darkening (Cassen {\it et al.}, 2006) which depends on
dimensionless coefficients $c_1$, $c_2$, wavelength $\lambda$,
projected distance between the center of the planet and the star
{\it r} and $R_{*}$. In this case the intensity is given by

\begin{equation}
I(\lambda,r) =
I(\lambda,0)[1-c_1(\lambda)(1-\mu)-c_2(\lambda)(1-\mu)^2],
\end{equation}

\noindent where
\begin{equation}
\mu = cos \theta = \sqrt{1-(r/R_*)^2}
\end{equation}
$\theta$ is the angle from the normal to the surface to the line of sight
and $I_{\lambda}(0)$ is the intensity emerging from the stellar disk center.

\noindent The flux from an unocculted star at the observer is
\begin{equation}
F \approx \int I d\omega
\end{equation}
where $\omega$ is the space angle subtended
by the star on the sky and $d\omega$ is given by
\begin{equation}
d\omega=2\pi \frac{R^2}{D^2}\sin \theta \cos \theta d\theta
\end{equation}
Then
\begin{equation}
F = \frac{2 \pi R^2}{D^2} \,I_{0} \int_0^{\pi/2}
[1-c_1(1-cos\theta)-c_2(1-cos\theta)^2] \,sin\theta\, cos\theta\,
d\theta
\end{equation}

\noindent where {\it D} is the distance between the observer and the
star. The detected flux of unocculted stellar disk leads to
\begin{equation}
F =
\pi\frac{R^2}{D^2}I_0\left[1-\frac{c_1}{3}-\frac{c_2}{6}\right]
\end{equation}

\noindent Assuming that $R_P \ll R$ the intensity can be
considered constant within the eclipsed region. Consequently,
within the 2-nd and 3-rd contact we may approximate the eclipsed flux
as $\delta F \approx I(\mu)\frac{\pi R_P^2}{D^2}$ and derive the normalized
lightcurve as

\begin{equation}
\frac{F-\delta F}{F}=1-\frac{\delta F}{F} = 1 -
\frac{1-c_1(1-\mu)-c_2(1-\mu)^2}{1-\frac{c_1}{3}-\frac{c_2}{6}}
\frac{R_P^2}{R^2}
\end{equation}

\noindent where $r = a \sqrt{sin^2(2 \pi t/P) + cos^2i cos^2(2\pi
t/P)}$. Eq.(7) is time dependent. We derived the
stellar limb darkening coefficients $c_1=0.38$ and $c_2=0.25$
according to the spectral type, mass and temperature of the parent
star COROT-Exo-2 (Table 1) from Claret {\it et al.}, 1995. Now, applying the nonlinear regression method on Eq.(7) we could
find its free parameters {\it $R_P$} and {\it i}. The method fits
the data in order to minimize the sum of $\chi^2$ (standard
deviations between the regression function and data points),
considering that we know the remaining variables of Eq.(7). Derived
parameters are shown in Table 1 and the fitting lightcurve with
measured data points in Figure 1. The magnitude drop also yields
from a model lightcurve.

\begin{table}[t]
\small
\begin{center}
\caption{Table of published transit epochs for COROT-Exo-2b, according to Exoplanet Transit Database, 2008.} \label{t2}
\begin{tabular}{lll}
\hline\hline
transit epoch (HJD) &  O - C (min) &  reference \\
\hline
2454237.53562 $\pm$ 0.00014  &   0   & Alonso {\it et al.}, 2008\\
2454624.4799 $\pm$ 0.0016    &  -0.0009  & Kleidis (AXA), 2008      \\
2454624.4831 $\pm$ 0.0011   &  0.0023  &   Ayiomamitis (AXA), 2008 \\
2454676.7577 $\pm$ 0.001    &  -0.013  &  Roe (AXA), 2008 \\
2454676.7612 $\pm$ 0.0008    &  -0.0095  &   Roe (AXA), 2008 \\
2454678.5016 $\pm$ 0.0012   &  -0.0121  & Mendez (AXA), 2008 \\
2454683.7403 $\pm$ 0.0014   &  -0.0024  & Roe (AXA), 2008   \\
2454685.4843 $\pm$ 0.0007    &  -0.0014  & Naves (AXA) \\
2454692.4595 $\pm$ 0.0015    & 0.0018   & Mendez (AXA), 2008 \\
2454697.6913 $\pm$ 0.0015    & 0.0046   & Roe (AXA), 2008    \\
2454702.9164 $\pm$ 0.0015    & 0.0007  & Foote (AXA), 2008    \\
\hline
2454706.4041 $\pm$ 0.003    &  0.0025 & Vere\v{s} {\it et al.}, 2008     \\
 \hline\hline
\end{tabular}
\end{center}
\end{table}

\section{Conclusions}
We carried out transit observations of the COROT-EXO-2b and a simple
analysis of the light curve. The transit depth in the R-filter was
$0.030\pm0.007\,mag$, which corresponds to the radius of the
exoplanet $R_P=1.318\pm0.158\,R_J$ and the inclination of the
orbit $i=87.88\pm0.15$. This is in agreement with the parameters
from the COROT mission. Our results confirm a relatively large
radius of the planet and evolutionary calculations with an
alternate source of energy such as a tidal heating might be needed
to explain it (see Burrow et al. 2008). The center of the transit
was observed at HJD $54706.4041\pm0.0030$ and the duration of the
transit was $2.24\pm0.15$ hours. In Figure 2 (see also Table
2) we put it into the context of other ground based observations
and study the O - C diagram. At present, we cannot confirm any
convincing evidence of the period variability which might have
been indicated by the three outlying points and our results are
in agreement with those of Alonso et al. (2008, 2008b).

Although the quality of the ground based data cannot be compared with
those from the COROT mission they still might help to constrain the properties
of the planet. Advantages of the ground observations are a long time baseline
and a wavelength limited spectral window. It is hoped that additional
ground based observations can improve the orbital period or reveal
possible changes in the planet's orbit. Also, additional
ground based observations in relatively well defined spectral regions
in the red part of the spectrum like those defined by the R, I filters
might have a good control over the limb darkening which
is wavelength dependent. Consequently, it might help to constrain
the planet's radius determined from the wide visual spectral region of COROT.
\\

\acknowledgements This work was supported by Vega Grant No.
$1/3067/06$ and grant for PhD students and postdocs Grant UK No.
$2008/399$ and Vega Grant No. 1/3074/06. J\'{a}n Budaj
acknowledges the support from the European International
Reintegration Grant MIRG-CT-2007-200297. We are thankful for
valuable comments and help to Juraj T\'{o}th, {\softL}ubom\'{i}r
Hamb\'{a}lek, Jean Schneider and for accessible actual data to the
Exoplanet Transit Database and The Extrasolar Planets
Encyclopaedia.



\end{document}